\documentstyle[sprocl,epsfig]{article}

\def\Journal#1#2#3#4{{#1} {\bf #2}, #3 (#4)}

\def\APJ{\emph{The Astrophysical Journal}}
\def\APS{\emph{The Astrophysical Journal Supplement Series}}
\def\ASJ{\emph{The Astronomical Journal}}
\def\MNR{\emph{Monthly Notices of the Royal Astronomical Society}}

\def\be{\begin{equation}}
\def\ee{\end{equation}}
\def\bea{\begin{eqnarray}}
\def\eea{\end{eqnarray}}

\begin{document}

\title{XMM--NEWTON OBSERVATION OF\\THE SEYFERT 2 GALAXY NGC4138}

\author{L. Foschini$^{(1)}$\footnote{Email: \texttt{foschini@bo.iasf.cnr.it}}, F. Panessa$^{(1)}$, A. L. Longinotti$^{(1)}$, L. C. Ho$^{(2)}$,
L. Bassani$^{(1)}$,\\
M. Cappi$^{(1)}$, M. Dadina$^{(1)}$, G. Malaguti$^{(1)}$, G. Di Cocco$^{(1)}$, F. Gianotti$^{(1)}$,\\
J.B. Stephen$^{(1)}$, M. Trifoglio$^{(1)}$}

\address{$^{(1)}$Istituto di Astrofisica Spaziale e Fisica Cosmica (IASF) del CNR\\Sezione di Bologna (formerly iTeSRE), Italy}
\address{$^{(2)}$The Observatories of the Carnegie Institution of Washington\\Pasadena (CA, USA)}


\maketitle\abstracts{The \emph{XMM--Newton} data presented here are the first X--ray
observation of the Seyfert 2 galaxy NGC4138. Although the galaxy has been pointed by ROSAT--HRI, it was not
detected, with a flux upper limit of about $1.1\times10^{-13}$~erg~cm$^{-2}$~s$^{-1}$
in the $0.2-2.4$~keV energy band~\cite{HAL}. \emph{XMM--Newton} performed the observation on 26 November 2001.
The source is detected for the first time in X--rays with $F_{0.2-2.4}=1.0\times 10^{-13}$~erg~cm$^{-2}$~s$^{-1}$,
in agreement with the upper limit of ROSAT~\cite{HAL}.
The source spectrum is typical of Seyfert 2 galaxies. We find heavy obscuration
($N_H\approx 8\times 10^{22}$~cm$^{-2}$) and a flat photon index ($\approx 1.6$). The source intrinsic luminosity
is about $5\times 10^{41}$~erg/s in the $0.5-10$~keV energy band.}

\section{Introduction}
NGC4138 is a spiral galaxy of Hubble type SA(r)0+ distant 17~Mpc, with a $D_{25}=2.57'$. Its morphology appears
undisturbed, although the galaxy has a dust lane in the southeastern side. The nucleus is classified as Seyfert 1.9~\cite{HO}.
Optical observations~\cite{JO} have shown evidence for a counterrotating disk, with a velocity dispersion
systematically lower than that of the primary disk. The counterrotating
disk may be either the result of a recent merger (it appears to
be still forming) or the continuum infall of material with opposite spin vector into
NGC4138. ROSAT--HRI observed the galaxy for $5.8$~ks without detecting it.
Halderson et al.~\cite{HAL} gave an upper limit to the $0.1-2.4$~keV flux of
$1.15\times 10^{-13}$~erg~cm$^{-2}$~s$^{-1}$. \emph{XMM--Newton} EPIC instrument detected
the source at a flux level of about $1.0-1.1\times 10^{-13}$~erg~cm$^{-2}$~s$^{-1}$ in the
same band.

\begin{figure}[h]
\begin{center}
\epsfig{figure=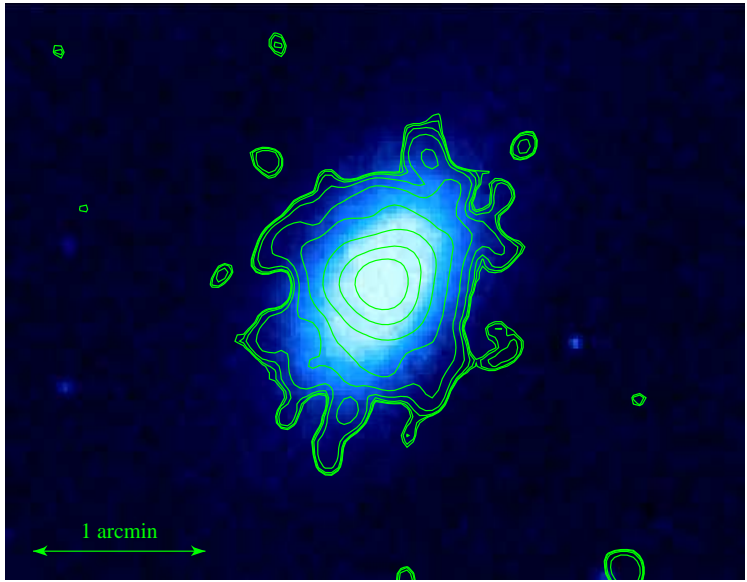, width=10cm}
\end{center}
\caption{Image from the Digitized Sky Survey with EPIC--MOS2 contours superimposed (energy band $0.5-10$~keV).
North is up and East is left. \label{images}}
\end{figure}

\begin{figure}[h]
\begin{center}
\epsfig{figure=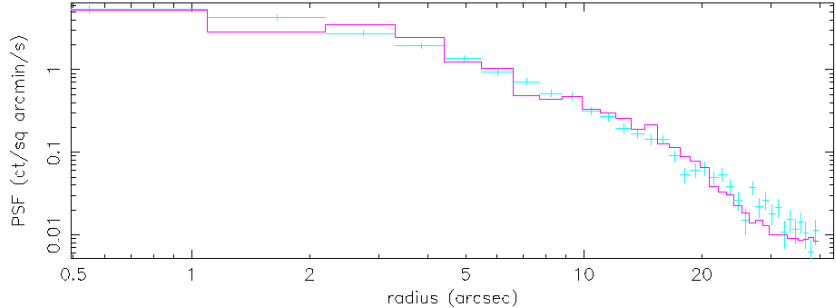, width=10cm}
\end{center}
\caption{Radial profile of NGC4138. Solid line: PSF at 5 keV;
Points: source counts in the energy range $0.5-10$~keV from EPIC--MOS2.\label{radial}}
\end{figure}

\begin{figure}
\begin{center}
\epsfig{figure=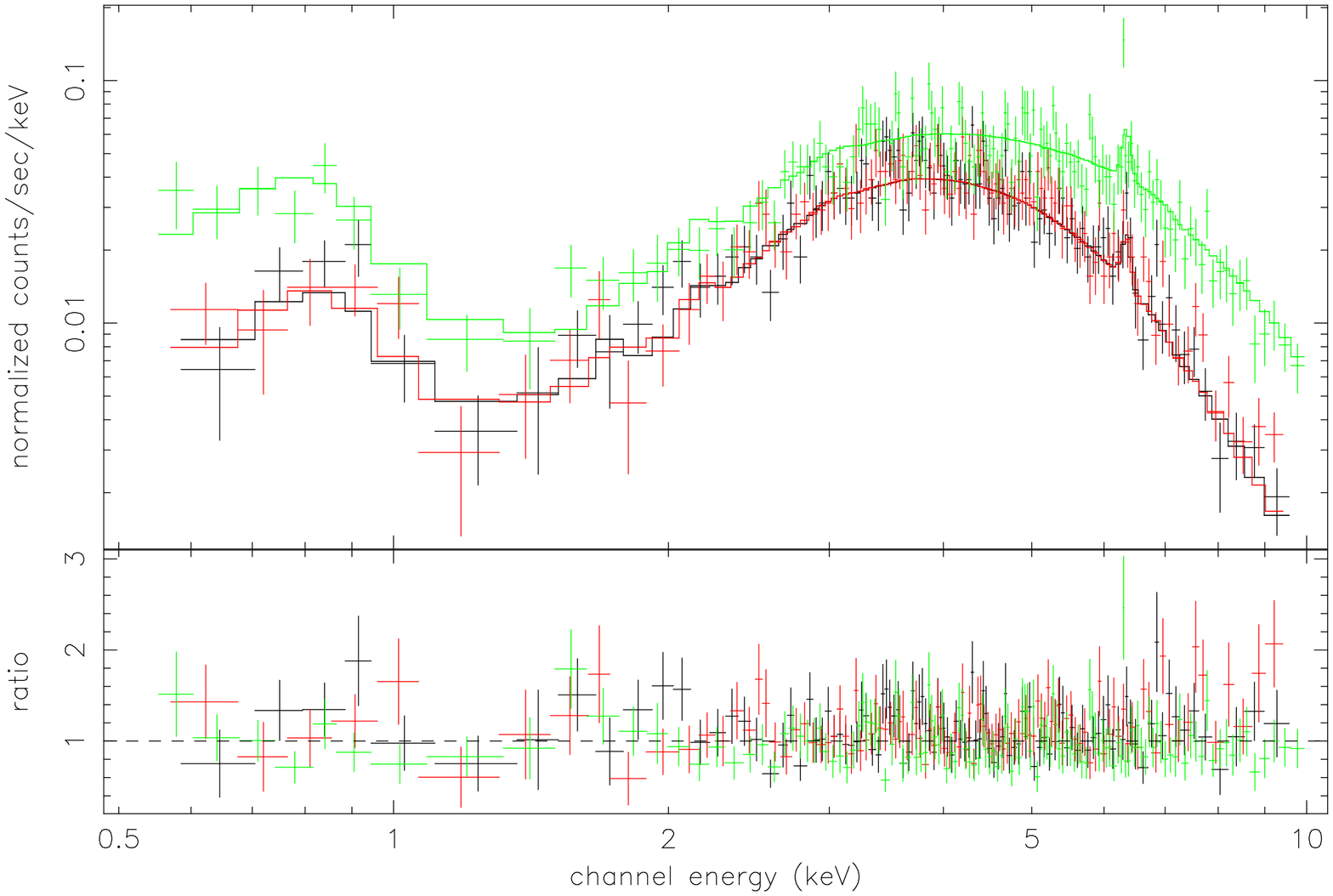, angle=270, width=10cm}
\end{center}
\caption{Best fit model for NGC4138: it consists of a power law plus
scattered component, and two gaussian emission lines at about $0.8$~keV and $6.4$~keV.
See text for more details. \label{spec}}
\end{figure}

\section{XMM--Newton data analysis}
In Fig.~\ref{images}, it is shown the Digitized Sky Survey (DSS) image of NGC4138, with
EPIC--MOS2 contours superimposed. Although optical observations indicate the presence of a
counterrotating disk, thus suggesting the possibility of a mild heating of the interstellar gas,
the \emph{XMM--Newton} data show no clear evidence of such process, in addition to that of the active nucleus.
The source appears to be point--like (Fig.~\ref{radial}). There is a marginal evidence of
a diffuse component beyond about $20''$ from the nucleus. A more detailed analysis will be performed soon.

The source spectrum is shown in Fig.~\ref{spec} and it is typical of a Seyfert 2 galaxy~\cite{TU}.
Indeed, its best fit model consists of an intrinsic absorbed power law component (with $\Gamma=1.6\pm 0.1$
and $N_H=(8.3\pm 0.7)\times 10^{22}$~cm$^{-2}$), plus a soft component dominating the emission below about
$1.5$~keV.

An Fe~K$\alpha$ line at $E=6.37\pm 0.04$~keV with $\sigma < 84$~eV and $EW\approx 100$~eV is also present.
This line is consistent with being produced by the transmission of X--rays through the measured absorption
column~\cite{LE}. It is hard to disentangle on the basis of the present data, if this absorber is the
putative torus of AGN or the dust lane visible in the optical observations~\cite{JO}.

The soft component is most likely due to the scattered primary radiation. In this case, it would account for
about 1\% of the direct emission. In this energy band, it is also present a broad line, or a blending of
non--resolved lines, at $0.79\pm 0.07$~keV (equivalent width $\approx 670$~eV). This feature suggests
the presence of either thermal and photoionization processes associated to a galactic hot gas.
It is possible to substitute the scattered power law and the emission line with the Mekal model of Xspec
for a hot plasma with emission lines. In this case, the temperature is $0.60\pm 0.09$~keV.

The relationships~\cite{DA} between the soft--X luminosity and the blue
and the FIR luminosity applied to the present case{\footnote{With the following values: observed
$L_{X,0.5-2~keV}=1.9\times 10^{39}$~erg/s, $L_{B}=1.5\times 10^{43}$~erg/s, $L_{FIR}=1.5\times 10^{41}$~erg/s.}}
suggest that the thermal emission may be due to hot gas derived from a star formation activity. However, the
optical observations of NGC4138~\cite{JO} showed that the star formation
activity in the primary disk ceased about 100~Myr ago. The only star formation activity present
in NGC4138 is in the counterrotating disk, especially in the area of a HII ring (with $22''$ radius).

More detailed observations with higher spatial resolution
(e.g., Chandra) are necessary to put stronger constraints on the nature of this hot gas and
on the structure of NGC4138.

\section*{Acknowledgments}
This work is based on observations obtained with \emph{XMM--Newton}, an ESA science
mission with instruments and contributions directly funded by ESA Member States and the
USA (NASA). This research has made use of the NASA's Astrophysics Data System Abstract Service and
of the NASA/IPAC Extragalactic Database (NED), which is operated by the Jet Propulsion Laboratory,
California Institute of Technology, under contract with the National Aeronautics and Space Administration.
We acknowledge the partial support of the Italian Space Agency (ASI) to this research.

\end{document}